\documentclass[conference, a4paper]{IEEEtran}

\usepackage{siunitx}
\usepackage{pbox}
\usepackage{makecell}
\usepackage{float}
 \usepackage{epsfig}
\usepackage[T1]{fontenc}
\usepackage[utf8]{inputenc}
\usepackage{graphicx}
\usepackage[square, numbers, comma, sort & compress]{natbib}
\usepackage{amsthm}      
\usepackage{amsmath}
\usepackage{url} 
\usepackage{amssymb}
\usepackage{caption}
\usepackage{subcaption}
\usepackage{acronym} 
\usepackage{mathrsfs}
\usepackage{csquotes}
\usepackage{booktabs}
\usepackage{url}
\usepackage{xcolor}
\usepackage[ruled,vlined,noend]{algorithm2e}
\usepackage[noend]{algpseudocode}
\usepackage{multirow}
\usepackage[shortcuts,acronym]{glossaries}
\usepackage{comment}
\usepackage{mathtools}
\usepackage{afterpage} 
\usepackage{pgf}
\usepackage{color}
\usepackage[normalem]{ulem}
\usepackage{balance}
\usepackage{svg}
\usepackage[utf8]{inputenc}
\usepackage{pgfplots}
\DeclareUnicodeCharacter{2212}{−}
\usepgfplotslibrary{groupplots,dateplot}
\usetikzlibrary{patterns,shapes.arrows}
\pgfplotsset{compat=newest}

\newacronym{g2g}{G2G}{Glass-to-Glass}
\newacronym{m2m}{M2M}{Motion-to-Motion}
\newacronym{e2e}{E2E}{End-to-End}
\newacronym{rpi}{RPi}{Raspberry Pi}
\newacronym{ntp}{NTP}{Network Time Protocol}
\newacronym{pps}{PPS}{Pulse Per Second}
\newacronym{gpio}{GPIO}{General-Purpose Input/Output}
\newacronym{cavs}{CAVs}{Connected and Autonomous Vehicles}
\newacronym{iqr}{IQR}{Interquartile Range}
\newacronym{nsa}{NSA}{Non-Standalone}
\newacronym{sa}{SA}{Standalone}
\newacronym{fov}{FOV}{field of view}
\newacronym{tod}{ToD}{Teleoperated driving}
\newacronym{spi}{SPI}{Serial Peripheral Interface}
\begin{document}
\title{End-to-End Latency Measurement Methodology for Connected and Autonomous Vehicle Teleoperation}
\author{\IEEEauthorblockN{Fran\c{c}ois PROVOST, Faisal HAWLADER, Mehdi TESTOURI and Raphaël FRANK}
\IEEEauthorblockA{Interdisciplinary Centre
for Security, Reliability and Trust (SnT)\\
University of Luxembourg, 29 Avenue J.F Kennedy,
L-1855 Luxembourg\\
firstname.lastname@uni.lu}
}
\maketitle
\begin{abstract}
Connected and Autonomous Vehicles (CAVs) continue to evolve rapidly, and system latency remains one of their most critical performance parameters, particularly when vehicles are operated remotely. Existing latency-assessment methodologies focus predominantly on Glass-to-Glass (G2G) latency, defined as the delay between an event occurring in the operational environment, its capture by a camera, and its subsequent display to the remote operator. However, G2G latency accounts for only one component of the total delay experienced by the driver. The complementary component, Motion-to-Motion (M2M) latency, represents the delay between the initiation of a control input by the remote driver and the corresponding physical actuation by the vehicle. Together, M2M and G2G constitute the overall End-to-End (E2E) latency. This paper introduces a measurement framework capable of quantifying M2M, G2G, and E2E latencies using gyroscopes, a phototransistor, and two GPS-synchronized Raspberry Pi 5 units. The system employs low-pass filtering and threshold-based detection to identify steering-wheel motion on both the remote operator and vehicle sides. An interrupt is generated when the phototransistor detects the activation of an LED positioned within the camera’s Field Of View (FOV). Initial measurements obtained from our teleoperated prototype vehicle over commercial 4G and 5G networks indicate an average E2E latency of approximately 500~ms (measurement precision ±4~ms). The M2M latency contributes up to 60\% of this value.
\end{abstract}

\vspace{3px}
\begin{IEEEkeywords}
Latency Measurement, Time Synchronization, Teleoperation, Connected and Autonomous Vehicles
\end{IEEEkeywords}
%

\begin{figure*}[!t]
    \centering
    \includegraphics[
        width=1\linewidth,
        height=7cm,
        keepaspectratio,
        trim=0cm 2cm 0cm 1.8cm,
        clip]{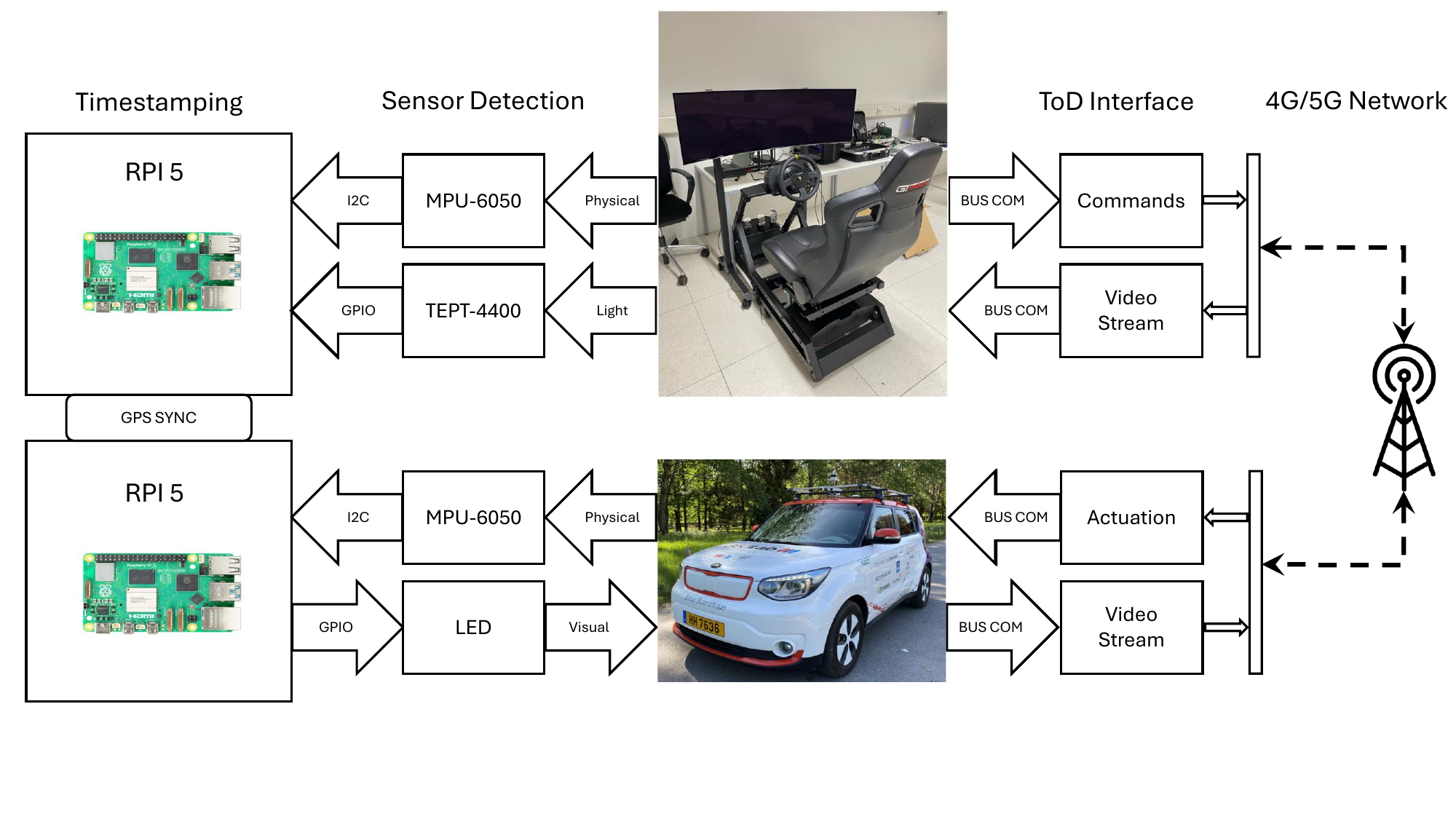}
    \caption{Measurement framework architecture, running on two GPS-synchronized \ac{rpi}~5, using two MPU-6050 for motion detection and TEPT4400 phototransistor for camera event detection.}
    \label{fig:m2m_latency_framework}
\end{figure*}

\vspace{-4px}
\section{Introduction}
\label{sec:introduction}
As highly Automated Driving Systems (ADS) continue to mature, teleoperation becomes a key enabler for their safe and scalable deployment. It provides a mechanism for remote supervision and intervention when the vehicle encounters complex or unforeseen situations beyond its operational design domain. Removing the operator from the vehicle is essential for making automated mobility services, such as robotaxis and automated public transport, commercially viable, as it allows a single operator to supervise multiple vehicles and reduces operational costs. To effectively perform this role, the underlying communication system must ensure a timely and precise interaction between the remote operator and the vehicle.

In such scenarios, the operator must be able to assume control without a perceptible delay. Minimizing the latency between the vehicle and the operator is, therefore, essential, as it directly affects both the perception of the environment and the execution of control commands. The study carried out in \cite{pantel_impact_2002} reported that performance started to degrade after latency reached 100~ms and became unacceptable after 500~ms, while \cite{chen_human_2007} recommended a maximum latency of 170~ms.

In order to mitigate the impact of latency, it is necessary to decompose the system into its individual contributing components and measure them. In teleoperation, latency affects both the perception of the environment transmitted to the operator, most commonly through a video stream, and the resulting actuation of the vehicle following the driver’s command. These are respectively referred to as \ac{g2g} and \ac{m2m} latencies. The sum of these two latencies is called \ac{e2e} latency.

Since latency is a critical factor in teleoperation, it has been extensively investigated in the literature, with various types of network being evaluated \cite{kaknjo_realtime_2018,uitto_evaluating_2022,yang_5g-nr_2022,den_ouden_design_2022}, particularly following the advent of 5G \ac{nsa} cellular networks and, in the future, 5G \ac{sa} with its very low-latency capabilities. Although latency measurement is widely reported, the methodologies used to obtain these measurements are rarely the primary focus of these studies. Existing research predominantly addresses \ac{g2g} latency \cite{bachhuber_system_2016,feldstein_simple_2021,ubik_video_2021}. To our knowledge, comprehensive studies specifically targeting the measurement of \ac{m2m} or even the full \ac{e2e} latencies have not yet been presented in detail.

Since \ac{m2m} latency has not yet been fully explored in the current literature, our previous work \cite{provost_motion--motion_2025} aimed to address this gap. However, the system had notable limitations. The Hall-effect sensor introduced calibration-dependent errors that exceeded 10~ms, and the method was highly sensitive to irregular steering-wheel motions, which delayed detection. The synchronization of the two \ac{rpi}~5 was also a limitation of our previous system. The use of a \ac{ntp} server, although averaging 300--400~µs, could increase to 4~ms, adding to the overall measurement error.

The system proposed in this paper\footnote{Source code available at: \url{https://github.com/sntubix/e2e_latency.git}}  aims to address these limitations. First, the Hall-effect sensor was replaced with gyroscopes that measure the angular velocity of the steering wheels. This enables detection of the onset of motion rather than relying on the wheel passing a single fixed sensor, thereby reducing sensitivity to sensor placement. The synchronization uses Chrony\footnote{Chrony: \url{https://chrony-project.org/}}
, but instead of a \ac{ntp} server, we use two GPS \ac{pps} receivers as sources to achieve more stable and accurate synchronization. The system also introduces \ac{g2g} latency measurement using the well-known LED and phototransistor solution described in \cite{bachhuber_system_2016}. A comprehensive experimental evaluation of our system, as installed, has been performed on our research vehicle \cite{testouri_5g-enabled_2025} together with its corresponding teleoperation interface \cite{testouri2025robocar}. In addition, an initial breakdown of the components of the measured latencies, including the latency induced by the network, is also provided.

\section{Related Work}
\label{sec:related_work}
\ac{tod} is widely considered a fallback strategy when automated driving systems reach their operational limits.
Prior work has shown that its feasibility and safety are strongly influenced by end-to-end latency \cite{9304802, 10588495}, since the remote operator must perceive the driving scene and execute corrective actions under strict timing constraints.
Human-factor studies further demonstrate that increasing end-to-end delay degrades operator performance and situational awareness \cite{munir2022situational}, thereby motivating stringent latency requirements for \ac{tod} \cite{chen_human_2007, pantel_impact_2002}.
To characterize delay sources, teleoperation latency has been decomposed into sensor-side (i.e., \ac{g2g}) \cite{feldstein2020simple} and actuator-side components (i.e., \ac{m2m}) \cite{black2024evaluation}, both of which can contribute substantially depending on the system pipeline and configuration \cite{georg_sensor_2020}.
Beyond the in-vehicle control loop, large-scale mobile-network measurements indicate that latency variability remains a key obstacle to reliable teleoperated driving, particularly under mobility \cite{neumeier_measuring_2019}. 
Similarly, end-to-end system designs over 4G and 5G highlight the coupled impact of network conditions and application-layer processing on \ac{tod} performance \cite{den_ouden_design_2022, kakkavas_teleoperated_2022}.

A commonly used metric to characterize the visual feedback chain is \ac{g2g} latency. 
For precise \ac{g2g} evaluation, \cite{bachhuber_system_2016} presented an optical method based on an LED phototransistor setup and demonstrated millisecond-level measurement precision.
Video-based measurement approaches have also been proposed to quantify end-to-end visual latency in teleoperation, achieving comparable measurement accuracy depending on the camera characteristics and experimental configuration \cite{feldstein_simple_2021, friston_measuring_2014}.
While these techniques provide high-fidelity characterization of the perception and display pipeline, they do not capture the command-to-actuation delay that determines the vehicle’s response time to operator inputs. 
To address this limitation, \ac{m2m} latency measurement has been proposed to quantify the command-to-actuation component of the teleoperation loop \cite{provost_motion--motion_2025, black2024evaluation}.
These evaluations often rely on application-level timestamps or network-layer measurements and typically do not instrument both ends with synchronized physical sensing to isolate \ac{m2m} latency precisely. 
In recent work, a motion-triggered \ac{m2m} measurement framework based on Hall-effect sensing \cite{provost_motion--motion_2025} was introduced. 
The framework exhibited calibration-dependent errors, sensitivity to irregular steering motion, and additional uncertainty due to clock-synchronization spikes, limiting repeatability at millisecond-level precision.
Related work has also discussed practical limitations of latency measurement when relying on external timing resources and highlighted the challenges of obtaining reliable measurements in \ac{tod} settings \cite{kaknjo_realtime_2018}.
However, this limitation can be mitigated through hardware-assisted synchronization \cite{buevich2013hardware, kyriakakis2018hardware}, which provides stable timing references for consistent correlation of events recorded at both the operator and vehicle sides.
This paper addresses this gap by presenting a hardware-synchronized measurement methodology that jointly quantifies \ac{g2g}, \ac{m2m}, and end-to-end latency, including baseline measurement precision and field validation across commercial 4G and 5G networks.

\section{System Architecture}
\label{Sec:Architecture}
\begin{figure*}[!t]
    \centering
    \includegraphics[width=0.98\linewidth,
        height=7cm,
        keepaspectratio,
        trim=0cm 7cm 0cm 6cm,
        clip]{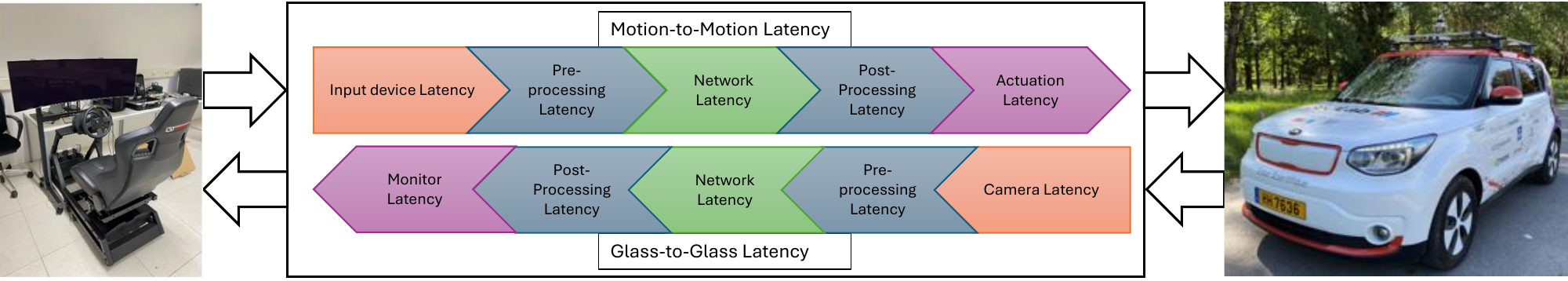}
    \caption{Relationships between the latency components and primary breakdown of \ac{m2m} and \ac{g2g}.}
    \label{fig:lat_break}
\end{figure*}

The overall architecture of the system can be found in Fig. \ref{fig:m2m_latency_framework}. Two gyroscopes are used, combined with software‑based low‑pass filtering and thresholding, to detect the motion of the steering-wheel from its onset to its completion. \(GY_{Station}\) and \(GY_{Vehicle}\) denote the timestamps corresponding to the onset of motion at the remote station and on the vehicle steering wheel, respectively. On the vehicle side, once the software detects a movement of the steering wheel, it activates an LED positioned in \ac{fov} of the camera, which is then detected by a phototransistor placed on the display screen. The delay between the detected onset of motion and the activation of the LED caused by thresholding is denoted as \(LED_{Delay}\). The actual LED activation time and the phototransistor activation time are indicated by \(LED_{On} \) and \(PT_{Trigger}\), respectively. The software logic is implemented in a custom Linux kernel module running on the \ac{rpi}~5 units, whose clocks are synchronized via GPS \ac{pps} through Chrony.
The latencies are then computed with the following formulas.

\begin{flalign}
M2M &= GY_{Vehicle} - GY_{Station} \label{eq:m2mform} \\
G2G &= PT_{Trigger} - LED_{On} \label{eq:g2gform} \\
E2E &= PT_{Trigger} - GY_{Station} - LED_{Delay} \nonumber\\
     &= PT_{Trigger} - GY_{Station} - (LED_{On} - GY_{Vehicle}) \nonumber\\
     &= M2M + G2G \label{eq:e2eform}
\end{flalign}

Each of these timestamped events can further be described by incorporating the effective error induced by the measurement system itself. In fact, all these values can be represented accordingly as follows:

\begin{flalign}
GY_{Vehicle}  &= M_{Vehicle}  + E_{GYv}  \label{eq:m2mvehicleerror}\\
GY_{Station}  &= M_{Station}  + E_{GYs}  \label{eq:m2mstationerror}\\
LED_{On}      &= LED_{Phy}    - E_{LED}  \label{eq:lederror}\\
PT_{Trigger}  &= PT_{Phy}     + E_{PT}   \label{eq:pterror}
\end{flalign}

\(M_{Vehicle}\) and \(M_{Station}\) denote the actual mechanical displacements of the respective steering wheels. \(E_{GYv}\) and \(E_{GYs}\) represent the delays between the physical movement of the wheel and the corresponding software-generated timestamped events. These delays are primarily determined by the sensor sampling rate, software-side data processing (including low-pass filtering and thresholding), and the Linux kernel scheduling latency.

\(LED_{On}\) denotes the software timestamp associated with the command to drive the LED. This event occurs before the LED is physically illuminated and subsequently detected by the phototransistor, denoted as \(LED_{Phy}\). The delay between these two events is represented by \(E_{LED}\). This delay depends mainly on the kernel scheduling latency and signal propagation time through the physical wiring.

\(PT_{Trigger}\) refers to the interrupt event generated when the phototransistor output transitions to a low state, while  \(PT_{Phy}\) denotes the corresponding physical detection event. The delay between these two events is defined as \(E_{PT}\). Similarly to \(E_{LED}\), this delay is primarily influenced by kernel scheduling and the signal propagation time within the circuit.

Substituting these delay components into the latency computation defined by formulas (\ref{eq:m2mform}), (\ref{eq:g2gform}), and (\ref{eq:e2eform}) yields the corresponding expression.

\begin{align}
M2M &= M_{Vehicle} + E_{GYv} - (M_{Station} + E_{GYs}) \nonumber\\
    &= M_{Vehicle} - M_{Station} + E_{GYv} - E_{GYs} \nonumber\\
    &= \underbrace{(M_{Vehicle} - M_{Station})}_{M2M_{Phy}}
     + \underbrace{(E_{GYv} - E_{GYs})}_{E_{M2M}} \nonumber\\
    &= M2M_{Phy} + E_{M2M}
\end{align}

\begin{align}
G2G &= PT_{Phy} + E_{PT} - (LED_{Phy} - E_{LED}) && \nonumber\\
    &= PT_{Phy} - LED_{Phy} + E_{PT} + E_{LED} && \nonumber\\
    &= \underbrace{(PT_{Phy} - LED_{Phy})}_{G2G_{Phy}}
     + \underbrace{(E_{PT} + E_{LED})}_{E_{G2G}} && \nonumber\\
    &= G2G_{Phy} + E_{G2G} &&
\label{eq:errg2gform}
\end{align}

\begin{align}
E2E &= M2M_{Phy} + E_{M2M} + (G2G_{Phy} + E_{G2G}) && \nonumber\\
    &= M2M_{Phy} + G2G_{Phy} + E_{M2M} + E_{G2G} && \nonumber\\
    &= \underbrace{(M2M_{Phy} + G2G_{Phy})}_{E2E_{Phy}}
     + \underbrace{(E_{M2M} + E_{G2G})}_{E_{E2E}} && \nonumber\\
    &= E2E_{Phy} + E_{E2E} &&
\label{eq:erre2eform}
\end{align}

where \(M2M_{Phy}\), \(G2G_{Phy}\), and \(E2E_{Phy}\) denote the corresponding true physical latencies, and \(E_{M2M}\), \(E_{G2G}\), and \(E_{E2E}\) represent the measurement-induced errors introduced by the system. These errors are further influenced by the accuracy of the clock synchronization. It should be noted that, in the case of \(E_{M2M}\), the hardware and software architectures deployed on both units are identical. Consequently, this latency component predominantly depends on how the low-pass filtering and thresholding algorithms process the motion signals on each side.

\subsection{Hardware}
The system uses two \ac{rpi}~5 units running Ubuntu~24.04. In our setup, movement detection is performed using two MPU‑6050 six‑axis IMUs mounted on each steering wheel, providing angular velocity measurements over the Inter-Integrated Circuit (I²C) protocol. A generic 3.3V green LED is detected on the display screen by a TEPT‑4400 phototransistor connected to a switching circuit that pulls the \ac{gpio} line to 0V when sufficient light is received. Chrony handles clock synchronization using \ac{pps} and NMEA signals provided by LC76G GNSS receiver modules, coupled with external Bingfu GPS antennas.

\begin{figure*}[t]
    \centering

    \begin{subfigure}[t]{0.45\textwidth}
        \centering
        \includegraphics[width=1\linewidth,
                     trim=0cm 0cm 0cm 0cm,
                     clip]{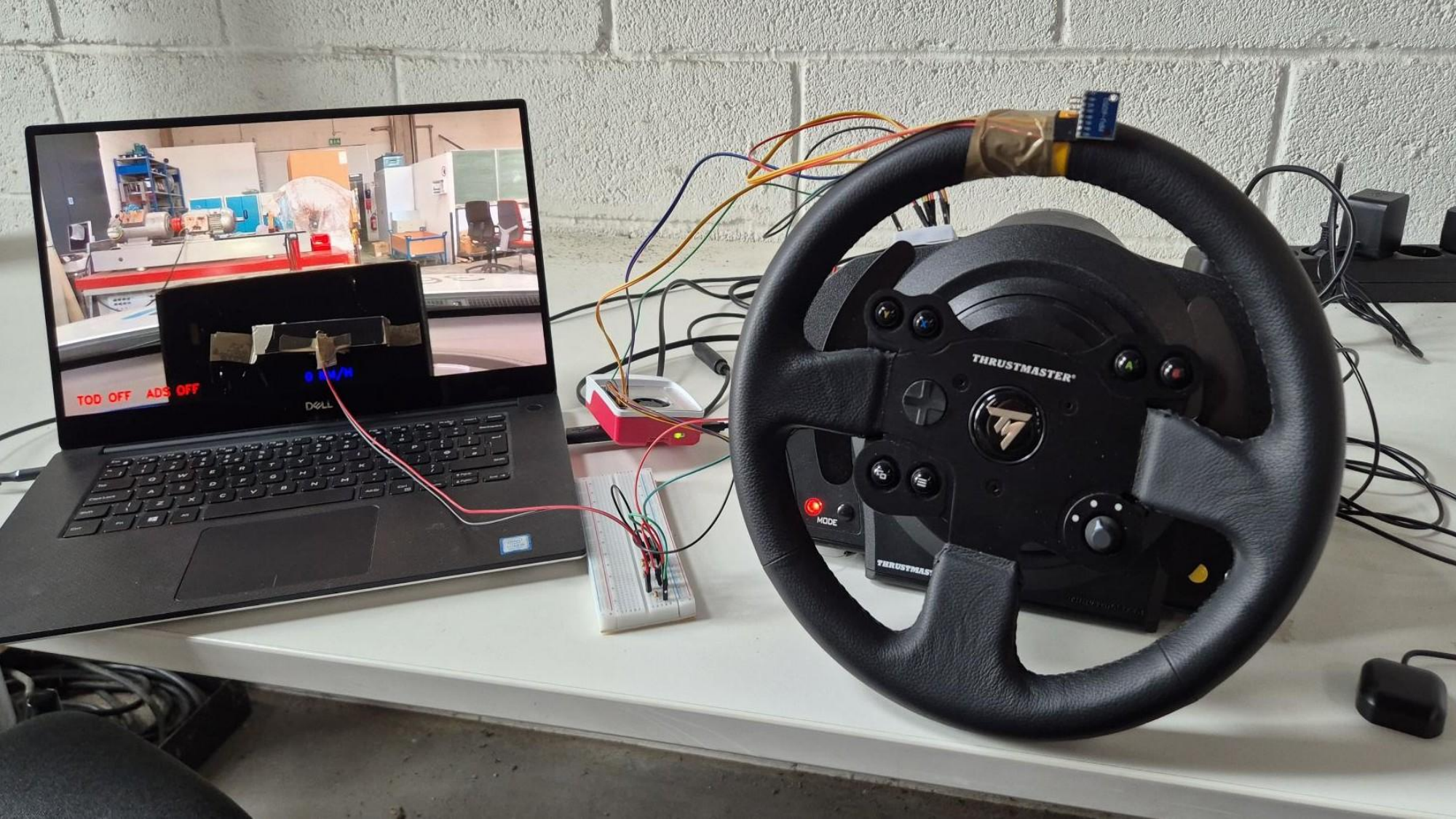}
        \caption{Station Setup}
        \label{fig:stationsetup}
    \end{subfigure}
    \hspace{0.5em} 
    \begin{subfigure}[t]{0.45\textwidth}
        \centering
        \includegraphics[width=1\linewidth,
                     trim=0cm 0cm 0cm 0cm,
                     clip]{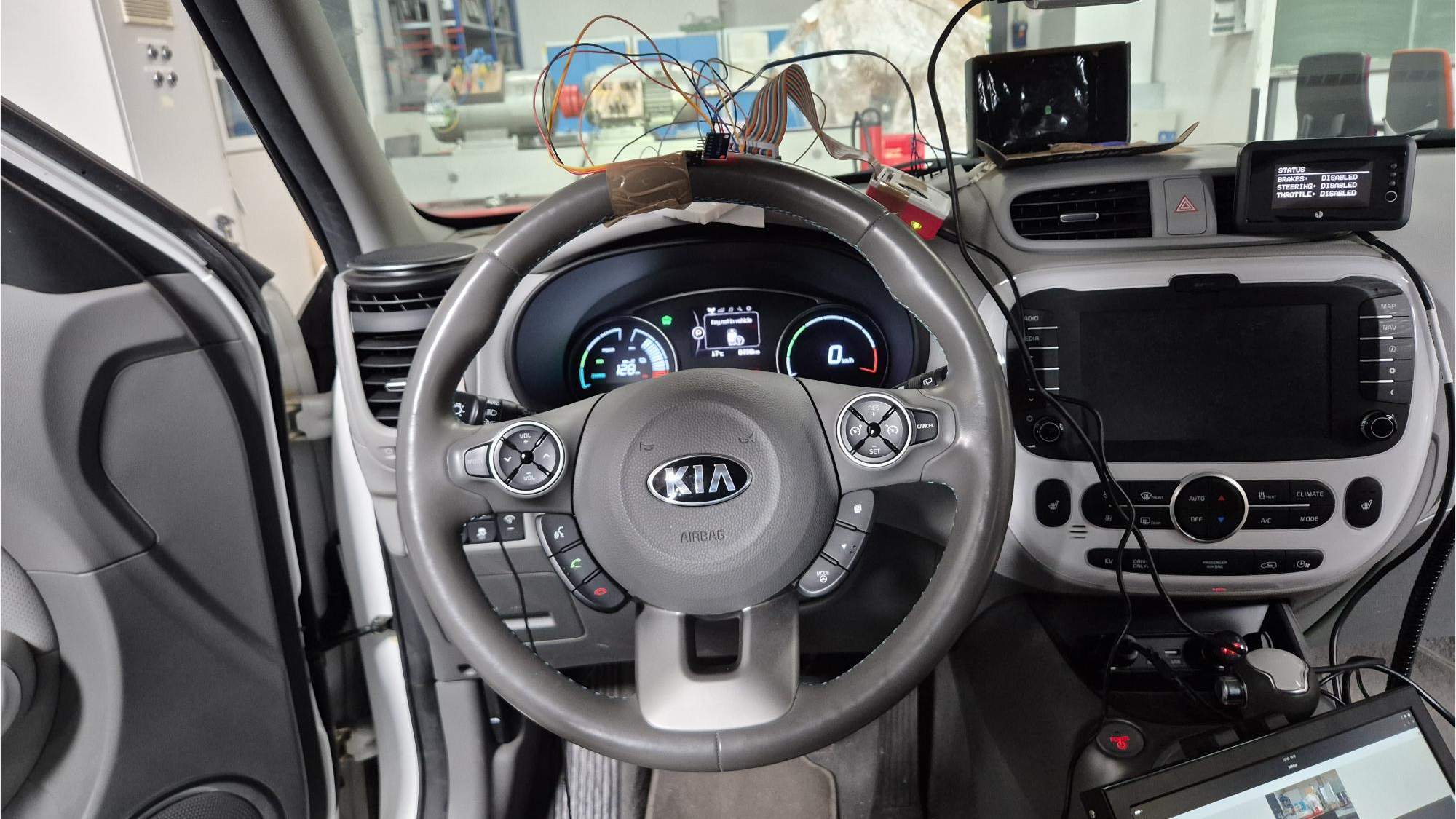}
        \caption{Vehicle Setup}
        \label{fig:vehiclesetup}
    \end{subfigure}
    \hspace{0.5em} 
\caption{Experimental setup for latency measurement framework field-tests.}
\label{fig:expsetup}
\end{figure*}

\subsection{Software}
The software logic runs in a custom kernel module that implements the interrupt routine for the \ac{gpio} connected to the phototransistor output, as well as a thread, woken by a high-resolution timer every 100~µs, that reads and processes the gyroscope data over I²C whenever possible. This timer interval was originally selected to accommodate the intended 400~kHz I²C communication rate. However, due to Linux‑related hardware initialization constraints, the effective rate was limited in practice to 97.5~kHz. This resulted in an average software‑loop duration of approximately 250~µs, of which I²C acquisition alone accounts for approximately 240~µs. Consequently, successive iterations are executed almost back-to-back, with only a 1-2~µs idle gap determined by the operating system scheduler.

The tri‑axial angular velocities measured by the gyroscopes are fused to estimate the relative velocity of each steering wheel, after which a low‑pass filter is applied to attenuate the measurement jitter. Each time the estimated velocity exceeds the predefined threshold, the corresponding timestamp is recorded.
When the velocity falls below the threshold, a counter is initiated. If it remains below the specified threshold for approximately 2.5 seconds, the motion is considered complete. This value was calibrated using the field-test setup described in Section~\ref{Sec:systeval}. The software also monitors the number of phototransistor detections that occur during a single LED activation and only considers measurements that generate exactly one detection event, thereby improving the reliability and consistency of the results.
This approach was designed to mitigate irregularities in the motion of the steering wheel of the vehicle that could otherwise result in incorrect detections.

Chrony handles clock synchronization, which is configured to use GPS NMEA data and the \ac{pps} signal instead of a standard NTP server.
This setup enables accurate board synchronization without any network connection, provided that a valid GPS signal is available.

\subsection{Latency breakdown}
Fig. \ref{fig:lat_break} presents a primary breakdown of the latencies that act between the remote operator and the vehicle. 
For the \ac{m2m}, the input device latency corresponds to the transformation of the mechanical movement of the steering wheel into digital position data transmitted via USB. These data are subsequently transmitted over the network (Network Latency) and undergo an initial processing stage (Pre-Processing Latency), during which the steering positions are converted into control commands for the vehicle. A second processing stage (Post-Processing Latency) follows before the command is delivered to the PID controller for actuation. Finally, vehicle actuation occurs under the control of the PID controller (Actuation Latency). It should be noted that the actuation latency also includes the mechanical delay between the input of the steering wheel and the corresponding movement of the vehicle’s wheels.

The \ac{g2g} follows a similar structural pipeline in which the data is transmitted to the remote station. The primary difference lies in the nature of the data being processed. In this case, frames of the environment are captured by the camera and transmitted via USB (Camera Latency). The frames are then processed by the stream server (Pre-Processing Latency), transmitted over the network (Network Latency), and subsequently processed by the stream client (Post-Processing Latency). Finally, the processed frames are displayed on the monitor (Monitor Latency).

A quantitative assessment of the contributions of these components is provided in Section~\ref{Sec:systeval}.

\section{System Evaluation}
\label{Sec:systeval}
This section presents the tests conducted to assess the reliability of the designed system for teleoperation of \ac{cavs}. We begin by establishing the system baseline through an analysis of its individual error sources, complemented by experimental measurements conducted over commercial 4G and 5G networks in Luxembourg City, provided by POST.

\begin{table}[h]
\centering
\resizebox{0.95\linewidth}{!}{
\begin{tabular}{lcccc}
 & \multicolumn{4}{c}{\textbf{Statistics}} \\
\cmidrule(lr){1-5}
\textbf{Synchronization Offset (µs)} & Min & Max & Mean & Std \\
\midrule
Offset & 0.257 & 12.544 & 3.226 & 2.217 \\
\midrule
\textbf{Baseline Latencies (ms)} & \multicolumn{4}{c}{} \\
\midrule
M2M & 0.207 & 7.357 & 3.475 & 2.076 \\
G2G & 0.460 & 0.487 & 0.470 & 0.005 \\
E2E & 0.688 & 7.826 & 3.945 & 2.077 \\
\bottomrule
\end{tabular}
}
\caption{\small Synchronization offset and measured baseline latencies.}
\label{tab:Synchronization}
\end{table}

\subsection{Baseline Tests}
The objective of this section is to determine the values of \(E_{M2M}\), \(E_{G2G}\), and \(E_{E2E}\) introduced in Section~\ref{Sec:Architecture}, including the effect of synchronization. To achieve this, the experimental conditions were designed such that \(M2M_{Phy}\), \(G2G_{Phy}\), and \(E2E_{Phy}\) are as close to zero as possible.

To approximate \(M2M_{Phy}\approx 0\), both gyroscopes were mounted on the steering wheel of the remote station, thereby measuring the same mechanical motion. The phototransistor was positioned directly in front of the LED inside an isolated enclosure to eliminate ambient light. Under these conditions, \(G2G_{Phy}\) corresponds only to the light propagation time between the LED and the phototransistor, which is considered negligible at the spatial scale used. Consequently, by definition, \(E2E_{Phy}\) is also assumed to be zero.

The latencies measured under these conditions, namely \ac{m2m}, \ac{g2g}, and \ac{e2e}, therefore correspond to the system-induced errors \(E_{M2M}\), \(E_{G2G}\), and \(E_{E2E}\), respectively. The results are summarized in Table~\ref{tab:Synchronization}.

In the same table, the synchronization accuracy and kernel scheduling latency are also reported separately. Note that these latencies are also components of the measured system-induced errors but are individually evaluated to highlight their potential influence. Synchronization accuracy was assessed by comparing the timestamps generated on each RPi 5 unit’s GPIO interrupt in response to the same physical event. The results of all the tests mentioned above are also presented in Table~\ref{tab:Synchronization}.

We observe that the synchronization error remains highly stable, confirming that both tests produce consistent results, with an average offset of 3~µs, a standard deviation of 2~µs, and a peak offset of approximately 12~µs. These results demonstrate that, provided that a stable GPS signal is available, synchronizing the boards with Chrony using GPS is more than sufficient for cross-platform latency measurements.

The measured \ac{g2g} latency is very stable with values between 460~µs and 480~µs and a standard deviation of 5~µs. In contrast, the \ac{m2m} latency exhibits a maximum offset of approximately 7~ms, with mean values around 3.5~ms. This indicates that, even though the hardware and software components are identical on both sides, each side processes the same motion differently. As a result, careful calibration of the low‑pass filter and detection thresholds is required to balance the detection speed while keeping false detections to a minimum. As noted above, \ac{e2e} latency corresponds to the sum of \ac{m2m} and \ac{g2g} latencies.

The kernel scheduling latency was measured using the Cyclictest utility from the Linux RT-Tests tool suite \footnote{RT-Tests: \url{https://documentation.ubuntu.com/real-time/latest/reference/real-time-metrics-tools/}}. The results indicate an average scheduling latency of 5~µs, with occasional outliers reaching up to 100~µs. These findings suggest that kernel scheduling latency affects system precision only sporadically and does not significantly impact overall timing performance.
\subsection{Field-tests}
\subsubsection{\textbf{Setup}}

The real-world capability of the system was evaluated using the RoboCar \cite{testouri2025robocar} TOD interface and vehicle configuration described in \cite{testouri_5g-enabled_2025}. The gyroscopes were installed on both the remote and vehicle steering-wheel, the LED was positioned directly in front of the camera within an isolated enclosure to minimize exposure to ambient light. The phototransistor was mounted to the monitor that displays the live camera feed.
All tests were carried out in a parking lot located in Kirchberg in Luxembourg\footnote{Test route: \url{http://g-o.lu/3/WYZn}}. 

Due to the fact that the gyroscopes and \ac{rpi}~5 units are directly connected via cables, the driving performance of the vehicle was constrained, making dynamic tests challenging. Consequently, measurements were conducted while the vehicle remained stationary in the parking lot to ensure an unobstructed GPS signal. The remote station was positioned in a nearby building with a clear view of the sky. The remote steering wheel was then operated and the corresponding timestamps were recorded.

The tests were conducted on commercial 4G and 5G non-standalone (NSA) networks, with more than 100 measurements collected for each experiment.

The experimental setup is illustrated in Fig. \ref{fig:vehiclesetup}.

\begin{figure*}[t]
    \centering
    \begin{subfigure}[t]{0.28\textwidth}
        \centering
        \includegraphics[height=0.19\textheight,keepaspectratio]{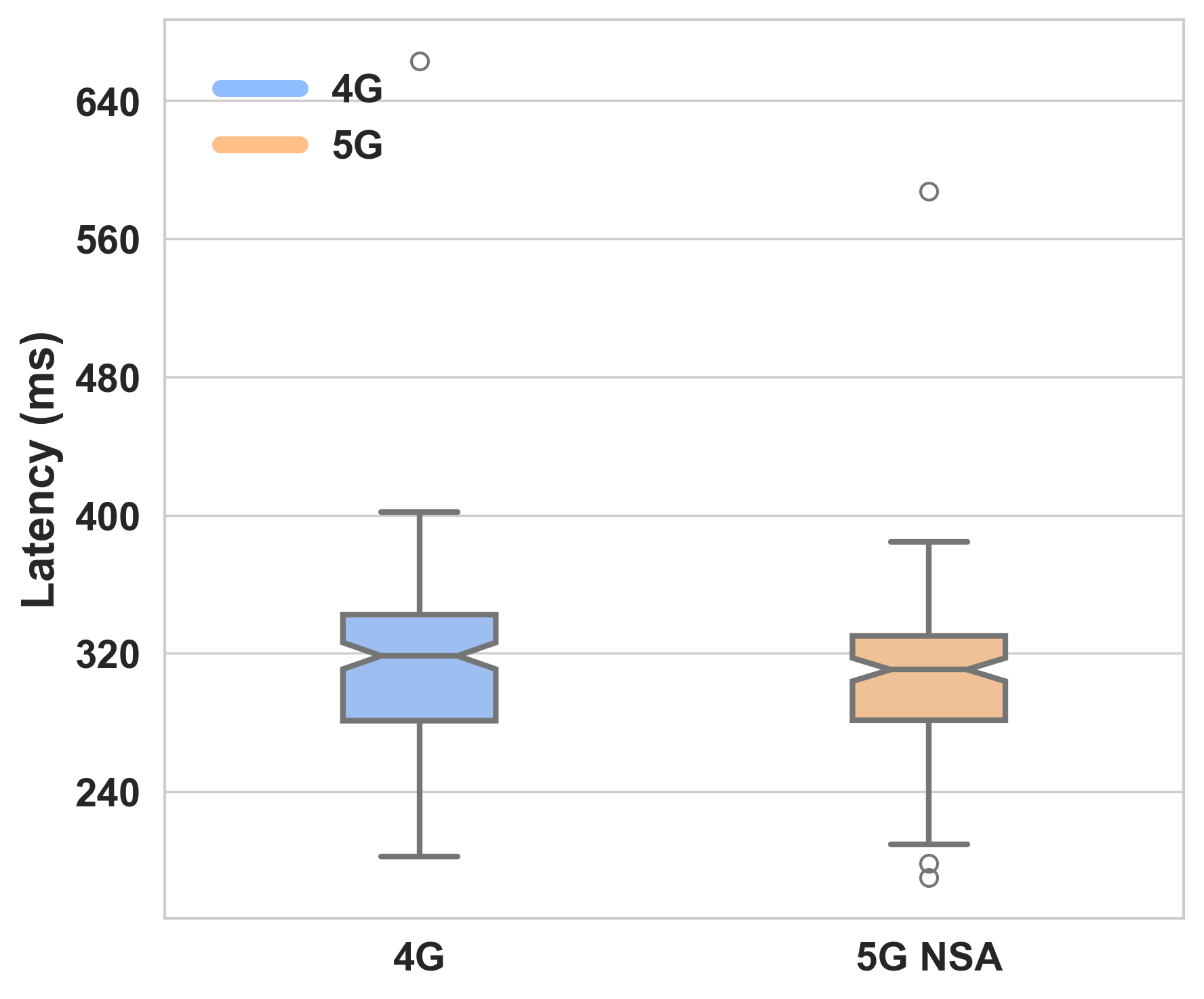}
        \caption{Motion-to-Motion (M2M)}
        \label{fig:m2m}
    \end{subfigure}
    \hspace{0.5em} 
    \begin{subfigure}[t]{0.28\textwidth}
        \centering
        \includegraphics[height=0.19\textheight,keepaspectratio]{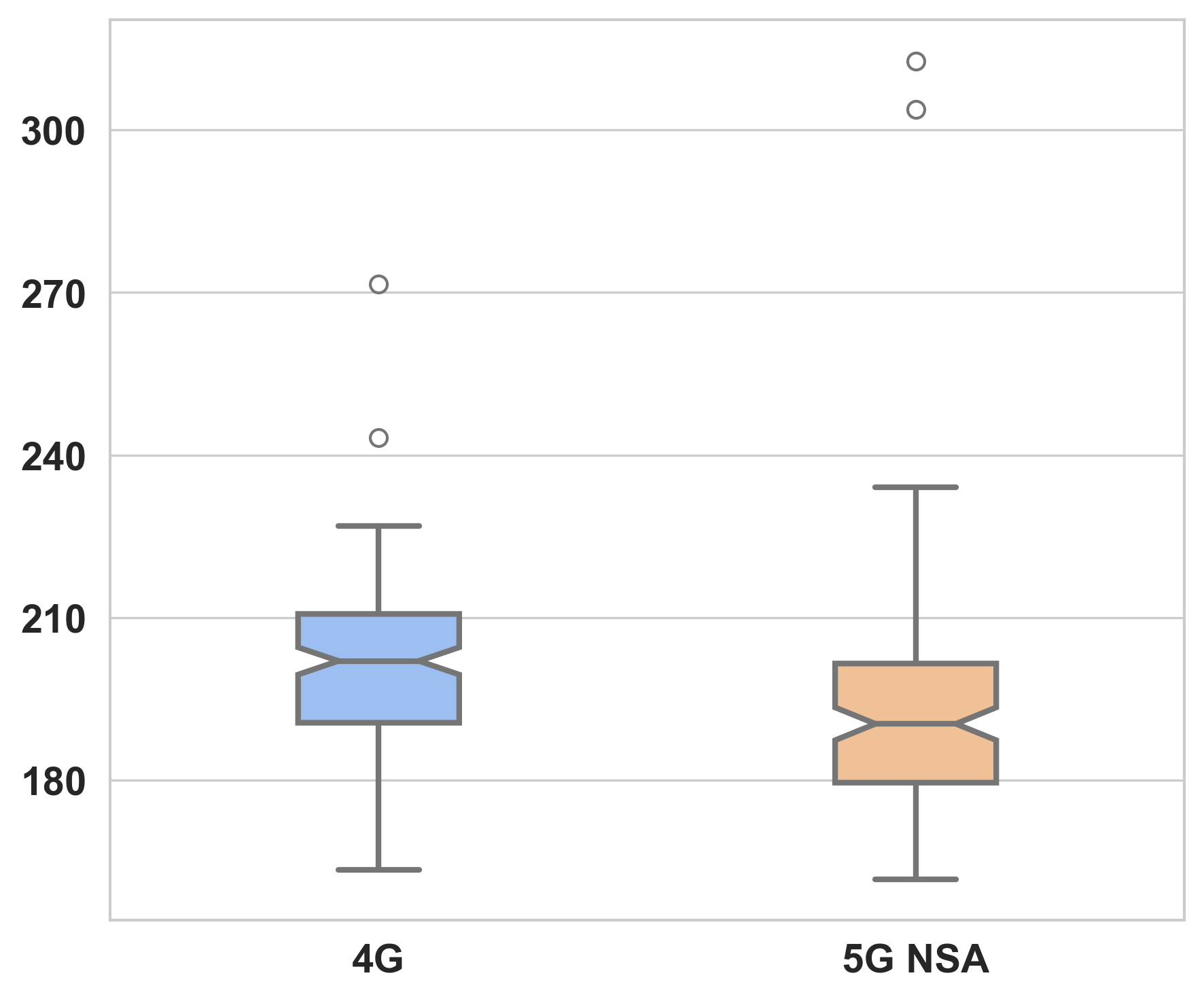}
        \caption{Glass-to-Glass (G2G)}
        \label{fig:g2g}
    \end{subfigure}
    \hspace{0.5em} 
    \begin{subfigure}[t]{0.28\textwidth}
        \centering
        \includegraphics[height=0.19\textheight,keepaspectratio]{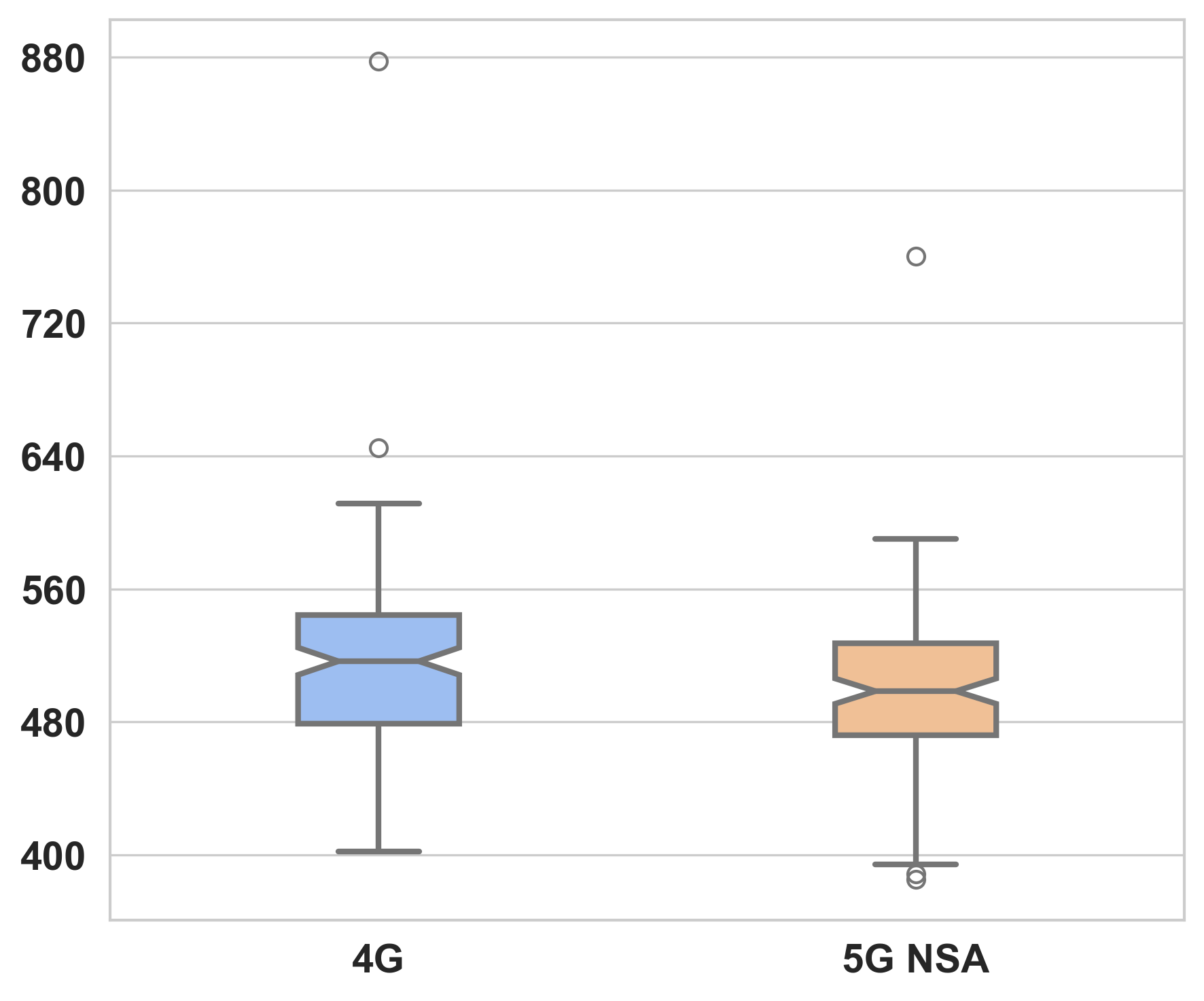}
        \caption{End-to-End (E2E)}
        \label{fig:e2e}
    \end{subfigure}

    \caption{\ac{m2m}, \ac{g2g}, and \ac{e2e} latency measurements using a 4G network compared with a 5G~\ac{nsa} network in a static scenario.}
    \label{fig:all_latency}
\end{figure*}

\subsubsection{\textbf{Results}}
The results presented in Fig. \ref{fig:all_latency} show that within the test area, 4G and 5G \ac{nsa} deliver broadly similar results across all latencies. This outcome is expected, as the 4G and 5G NSA networks inherently share the same core network, limiting the capabilities of 5G \ac{nsa} compared to standalone 5G. For 4G, the median \ac{m2m} latency is 318~ms with an \ac{iqr} of 61~ms, while the \ac{g2g} latency exhibits a median of 202~ms and a \ac{iqr} of 20~ms. The combined \ac{e2e} latency reaches a median of 516~ms with an \ac{iqr} of 65~ms. 
In comparison, 5G \ac{nsa} reports a median \ac{m2m} latency of 311~ms (\ac{iqr}: 49~ms), a median \ac{g2g} latency of 190~ms (\ac{iqr}: 22~ms) and a median \ac{e2e} latency of 498~ms (\ac{iqr}: 56~ms).

In general, these results indicate a modest reduction in median latency with 5G \ac{nsa}, both in absolute latency and stability, with reductions that generally range from 10 to 20~ms while variability remains comparable.

The \ac{m2m} latency measurements show a clear improvement compared to the results previously reported in~\cite{provost_motion--motion_2025}, where the median latency for both the 4G and 5G tests was close to 800~ms. This discrepancy can be explained by differences in the way the onset of motion was detected. In the earlier work, the measurement was triggered when the magnet passed in front of the hall-effect sensor. Consequently, any irregularities in motion caused by the PID controller could potentially add a delay to the final measurement.

\subsection{Latency Breakdown}
\begin{figure}[t]
    \centering
    \begin{subfigure}[t]{0.95\linewidth}
        \centering
        \includegraphics[width=1\linewidth]{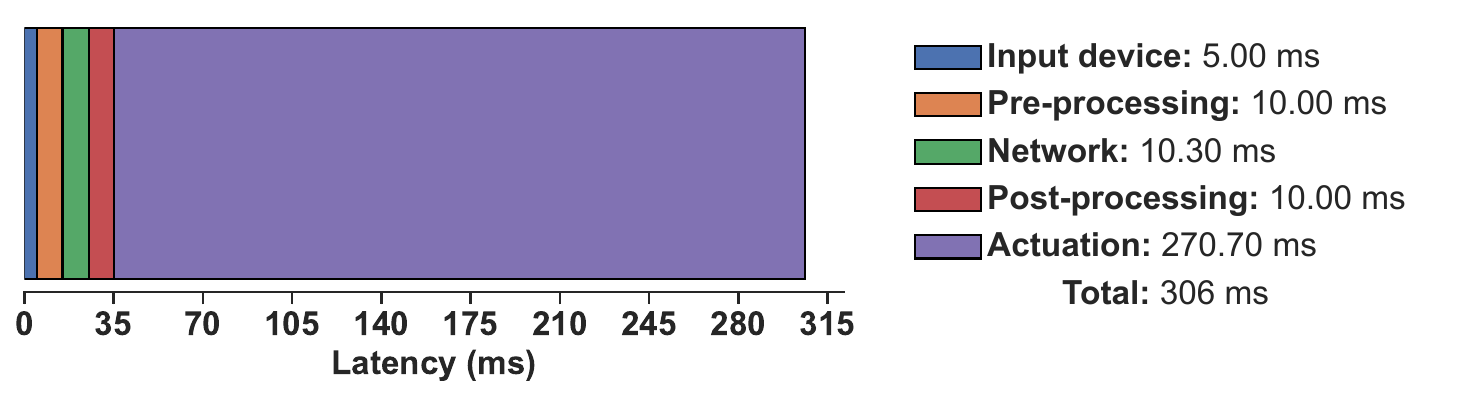}
        \caption{Motion-to-Motion (M2M)}
        \label{fig:m2m_results}
    \end{subfigure}

    \vspace{0.5em} 

    \begin{subfigure}[t]{0.95\linewidth}
        \centering
        \includegraphics[width=1\linewidth]{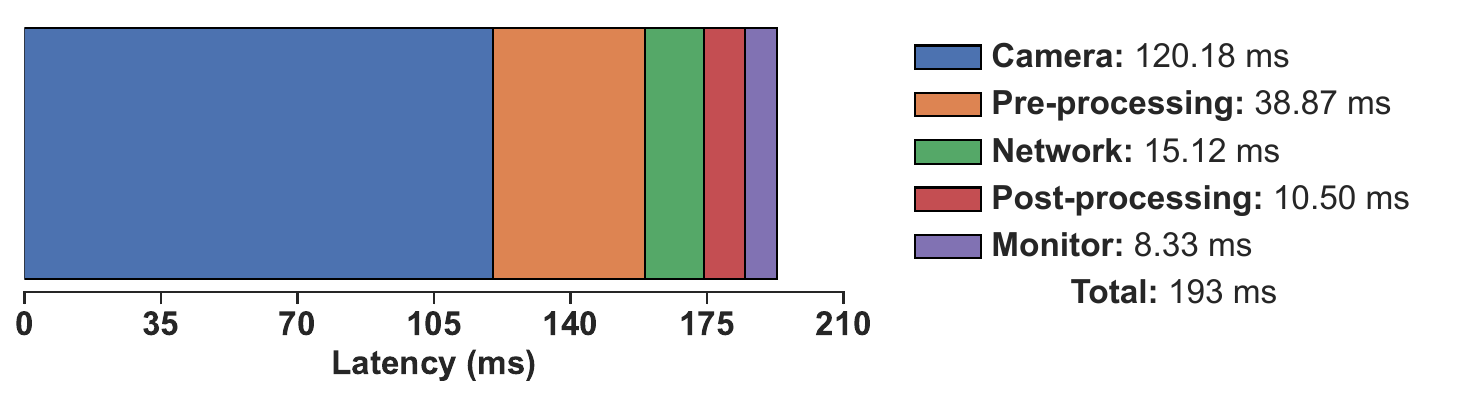}
        \caption{Glass-to-Glass (G2G)}
        \label{fig:g2g_results}
    \end{subfigure}

    \caption{Motion-to-Motion and Glass-to-Glass breakdown using the 5G \ac{nsa}  network.}
    \label{fig:breakdownresults}
\end{figure}

For this primary breakdown of latency we chose to focus on the 5G \ac{nsa} network expecting 4G network to provide close results in the conducted field-tests. The mean \ac{m2m}, \ac{g2g} and \ac{e2e} latencies from the 5G \ac{nsa} field test dataset, respectively, 306~ms, 193~ms, and 499~ms, are used as reference values for components that could not be measured directly. The contribution of the uncharacterized components is estimated, for each of the \ac{m2m} and \ac{g2g} segments, as the difference between the corresponding total latency and the sum of the directly measured components. A summary of the breakdown is presented in Fig.~\ref{fig:breakdownresults}.
\vspace{0.5em}

\textbf{Motion-to-Motion Latency:}
As defined in Fig. \ref{fig:lat_break}, \ac{m2m} latency can be decomposed into the input device latency, which in our case corresponds to the delay between the mechanical movement and the transmission of its position data via USB. Although the manufacturer does not provide official specifications for this latency, it was estimated by capturing the USB packets using Wireshark. With this method, we compute the average polling interval to be approximately 2~ms, accepting some error margin in case position is sent twice, it would bring the input device latency around 5~ms.

The \ac{tod} interface operates at a frequency of 100 Hz on both the remote station and the vehicle sides, meaning that commands are processed and transmitted every 10~ms. This interval represents our command pre-processing and post-processing latency.

The network latency was evaluated by capturing and analyzing the command TCP packets in Wireshark, yielding an average latency of 10.30~ms.

The actuation latency could not be measured directly. It was estimated based on the average measured \ac{m2m} latency (306~ms) from field tests, subtracting the contributions of the other components, resulting in an estimated actuation latency of approximately 270~ms, which includes the PID controller sending the command, including the delay introduced by the PID controller, mechanical actuation of the wheels, and the resulting steering wheel motion.
\vspace{0.5em}

\textbf{Glass-to-Glass Latency:}
Regarding \ac{g2g} latency, camera frames are published through a ROS2 publisher at a period of 30~ms. The streaming server loop subscribes to this publisher and is triggered whenever new data become available, adding its execution time to the pre-processing latency. This execution time was monitored using timestamp measurements and exhibited an average value of 8.87~ms, resulting in a total pre-processing latency of approximately 39~ms.

To estimate network latency, Wireshark was used to capture and analyze the UDP packets transmitted by the video stream, resulting in an observed application-layer throughput of 1376 kb/s, excluding protocol overhead. Concurrently, the average compressed image size was measured at 20.8 kb, with observed values ranging from 10 to 50 kb. The latency breakdown presented in this study is based on the mean image size. Using these measurements, the network latency was determined to be approximately 15~ms.

On the streaming client side, processing is performed within a callback that is triggered whenever data become available from the pipeline. 
Consequently, only the callback execution time contributes to the total latency. Similarly to the execution time on the server-side, this duration was measured using timestamping and exhibited an average value of 10.5~ms.

The monitor displaying the video feed operates at 60 Hz, implying that a new frame can be presented within an interval ranging from 0 to 16.67~ms. Assuming a uniform distribution of this component, the corresponding average display latency is approximately 8.33~ms.

The remaining 120~ms can therefore be attributed to camera latency, defined as the delay between the physical event and the delivery of the corresponding frame to the host system via USB. Although direct measurements of the latency of this particular camera are not available, the estimated value appears consistent with independent \ac{g2g} tests performed using a similar device directly connected to a computer via USB, which observed minimum latencies of approximately 100~ms.

Since \ac{e2e} latency represents the cumulative delay of both the \ac{m2m} and \ac{g2g} components, a more detailed breakdown is not particularly meaningful. Nevertheless, it can be observed that, for both 4G and 5G field tests, \ac{m2m} accounts for approximately 60\% of total latency on average, while \ac{g2g} contributes roughly 40\%.

At the component level, camera latency and vehicle actuation together account for nearly 400~ms of the total of 500~ms \ac{e2e} latency, highlighting their dominant contribution to overall delay.

\section{Conclusion and Future Work}

This paper presents a methodology for measuring \ac{m2m}, \ac{g2g}, and \ac{e2e} latencies in the remote operation of \ac{cavs}. The proposed measurement framework achieves baseline measurement precisions of 3.5~ms, 0.5~ms and 4~ms for \ac{m2m}, \ac{g2g}, and \ac{e2e}, respectively. These figures include a synchronization offset with an average value of 3~µs, demonstrating stable and highly accurate time alignment between measurement units.

Field tests revealed average \ac{m2m}, \ac{g2g}, and \ac{e2e} latencies of approximately 300~ms, 200~ms and 500~ms, respectively, under both 4G and 5G \ac{nsa} connectivity. Although 5G \ac{nsa} consistently exhibited slightly lower latencies, typically in the range of 10–20~ms, the overall latency magnitude remained comparable within the evaluated test area.

The breakdown of \ac{g2g} latency indicates that, while network transmission is not negligible, accounting for roughly 10\% of the total \ac{g2g} latency excluding encoding and decoding, hardware components can also constitute significant contributors if not carefully selected. In particular, low camera sampling rates and limited display refresh rates introduce additional delays that directly increase the measured latency. The transmission delay between the camera and the streaming interface may also represent a relevant contribution; however, it could not be isolated with the current experimental setup. These findings underline the importance of jointly monitoring \ac{m2m} and \ac{g2g} latencies and considering not only network performance, but also sensing, processing, and hardware-induced delays when optimizing the overall system.

Future work will address the current limitations of the framework. For instance, the error in the \ac{m2m} latency measurement could be reduced by hardware optimization, such as replacing the I²C interface with \ac{spi}. Introducing a wireless link between the gyroscopes and \ac{rpi}~5 units could further reduce the operational constraints imposed by the current wired configuration. 
Finally, a more granular latency decomposition could be achieved by extending the synchronization mechanism to additional intermediate processing components on both the vehicle and remote-station sides, allowing higher-precision timestamping of internal processing stages.

\section*{Acknowledgments}
This work is supported by the Luxembourg National Research Fund under the 5G BRIDGES/2023-Phase 2/IS/19101381/5GDrive project.
\balance 
\bibliographystyle{IEEEtran}

\end{document}